\documentclass[9pt]{article}
\usepackage{spconf,amsmath,graphicx}
\usepackage{hyperref}
\newcommand{\norm}[1]{\left\lVert#1\right\rVert}
\usepackage{booktabs}
\usepackage{xcolor,colortbl}
\usepackage{tikz}
\def\checkmark{\tikz\fill[scale=0.4](0,.35) -- (.25,0) -- (1,.7) -- (.25,.15) -- cycle;}

\title{Zero-Shot Multi-Speaker Text-To-Speech with State-of-the-art Neural Speaker Embeddings}

\name{Erica Cooper$^{\star}$, Cheng-I Lai$^{\dagger}$\thanks{The second author performed the work mostly while interning at NII.}, Yusuke Yasuda$^{\star}$, Fuming Fang$^{\star}$, Xin Wang$^{\star}$, Nanxin Chen$^{\ddagger}$, Junichi Yamagishi$^{\star}$}

			\address{$^{\star}$ National Institute of Informatics, Tokyo, Japan \\
			    $^{\dagger}$ Massachusetts Institute of Technology, Cambridge, USA 
			    $^{\ddagger}$ Johns Hopkins University, Baltimore, USA}

\begin{document}
\ninept

\maketitle

\begin{abstract}

While speaker adaptation for end-to-end speech synthesis using speaker embeddings can produce good speaker similarity for speakers seen during training, there remains a gap for zero-shot adaptation to unseen speakers.  We investigate multi-speaker modeling for end-to-end text-to-speech synthesis and study the effects of different types of state-of-the-art neural speaker embeddings on speaker similarity for unseen speakers.  Learnable dictionary encoding-based speaker embeddings with angular softmax loss can improve equal error rates over x-vectors in a speaker verification task; these embeddings also improve speaker similarity and naturalness for unseen speakers when used for zero-shot adaptation to new speakers in end-to-end speech synthesis.

\end{abstract}


\section{Introduction}
\label{sec:intro}

Recent advances in end-to-end text-to-speech (TTS) synthesis have enabled us to produce very realistic and natural-sounding synthetic speech \cite{wang2017tacotron,ping2018clarinet} with mean opinion scores (MOS) approaching those of natural human speech \cite{shen2018natural}. Not only speaker dependent TTS systems but also multi-speaker TTS systems show remarkable results \cite{abs-1907-04462}. However, adapting voice models to arbitrary new speakers using a small amount of data (speaker adaptation) remains a challenge.

An effective approach for speaker adaptation in neural TTS is to fine-tune all or part of model with a small amount of data from the target speaker \cite{kons2019high,chen2019sample}.  This approach can also be used to adapt to new speaking styles such as Lombard speech \cite{bollepalli2019lombard}.  A different but complementary approach is to use speaker embeddings to model speaker identity in TTS.  Prior studies have focused on training a speaker encoder network jointly with the TTS model \cite{chen2019sample,park2019multi,nachmani2018fitting} or the neural vocoder \cite{deng2018modeling}; others have explored the use of speaker embeddings in combination with fine-tuning the TTS model \cite{deng2018modeling,hu2019neural,arik2018neural}.  Approaches that use fine-tuning necessarily require transcribed adaptation data, as well as more computational time and resources to adapt to a new speaker.  Furthermore, speaker encoder networks that are jointly trained with the TTS model cannot benefit from data outside of the TTS training data, which is restricted to be of relatively high quality in clean recording conditions.  

Transfer learning for speaker modeling in TTS addresses these issues.  With this approach, the speaker embedding network is trained completely separately, perhaps for a different task such as speaker recognition.  The benefit of this approach is that speaker recognition models can be trained on a large amount of data that does not have to be of the same high quality typically required for TTS, and these models can obtain robust speaker representations that are independent of channel and recording conditions using relatively small amounts of target speaker data, which does not necessarily have to be transcribed.  End-to-end synthesis models can then be used to adapt to a target speaker's voice in a zero-shot manner by using the speaker embedding only, without necessarily needing to fine-tune the entire model.  Several recent studies \cite{nachmani2018fitting,jia2018transfer,pascual2019learning,chen2019cross} have used this approach for speaker modeling in TTS, with \cite{chen2019cross} modeling both speaker and language characteristics.  \cite{jia2018transfer} observed that unseen speakers' synthetic speech had lower speaker similarity to the target speaker than seen speakers, accents were often mismatched, and nuances such as characteristic prosody were lost, indicating that while seen speakers can be well-modeled in this manner, there is room for improvement for modeling unseen speakers.

In parallel with the above-mentioned studies, there has been substantial development in end-to-end speaker recognition. Villalba et al. summarized several state-of-the-art speaker recognition systems for the NIST SRE18 Challenge \cite{villalba2019state}, where x-vector based systems \cite{snyder2018x} consistently outperformed i-vector based systems \cite{dehak2010front}. There has also been a surge of interest in new encoding methods and end-to-end loss functions for speaker recognition \cite{cai2018exploring, Huang2018, chen2019tied, xie2019utterance, hajibabaei2018unified, xiang2019margin, jung2019spatial}. One prominent advancement is the use of learnable dictionary encoding (LDE) \cite{cai2018exploring} and angular softmax \cite{Huang2018} for speaker recognition, which are reported to boost the speaker recognition performance on open-source corpora such as the VoxCelebs \cite{nagrani2017voxceleb, chung2018voxceleb2}. 

One aspect of our study is therefore an attempt to find out how effective these recent developments in speaker verification are for speaker adaption in TTS.
More specifically we investigate the capability of neural speaker embeddings \cite{villalba2019state, snyder2018x, cai2018exploring} to capture and model  characteristics of speakers that were unseen during TTS model training. For this purpose, we extend an improved Tacotron system in \cite{yasuda2019investigation} to a multi-speaker TTS system and conduct systematic analysis to answer the above question. We also analyze how the quality and similarity of generated voices are correlated with automatic speaker verification (ASV) accuracy.

While prior studies have focused on transfer learning for zero-shot speaker adaptation for end-to-end TTS, to our knowledge this is the first investigation of many different types of speaker embeddings to determine whether some type of embedding is best for modeling unseen speakers and to learn whether the best embeddings for ASV are the same as the best embeddings for TTS.

\vspace{-2mm}        
\section{Neural Speaker Embeddings}
\label{sec:speaker_emb}
There are three components in a typical end-to-end speaker recognition system: an encoder network, a statistical pooling layer, and a classifier \cite{villalba2019state}. An encoder network acts as a frame-level feature extractor, the statistical pooling layer summarizes frame-level representations to a fixed-dimensional utterance-level embedding, and the classifier determines the speaker identity based on the embedding. In most cases, the neural speaker embedding is obtained after pooling and before classification. Below, we describe each of these components used in our work.  

\vspace{-2mm}
\subsection{Encoder Network}
In the original x-vector paper \cite{snyder2018x}, a time delay neural network (TDNN) was used as the encoder network and its variants were explored in \cite{villalba2019state}. The TDNN is composed of 1D convolution and fully connected layers. 
Several later studies \cite{villalba2019state, cai2018exploring, chen2019tied, xie2019utterance, jung2019spatial, chung2018voxceleb2} suggest replacing the TDNN with variants of ResNet34, composed of 2D convolutions, as the encoder network. We used TDNN and ResNet34 for x-vector and LDE embeddings, respectively. 
    
\vspace{-2mm}    
\subsection{Pooling methods}

The pooling method is an important component since it summarizes frame-level representations into a fixed-dimensional utterance-level embedding. 

\noindent\textbf{Statistical Pooling (SP)}: A statistical pooling layer was adopted in the original x-vector paper \cite{snyder2018x}. It computes the mean and standard deviation of the frame-level representations, which are concatenated as a fixed-dimensional vector. 
        
\noindent\textbf{Learnable Dictionary Encoding (LDE)}: Instead of the single mean and standard deviation as in SP, the LDE layer proposed in \cite{cai2018exploring} conducts soft clustering of the frame-level representations and concatenates the clusters' means and standard deviations. 

\begin{figure}[tb] \centering
    \includegraphics[width=0.7\columnwidth]{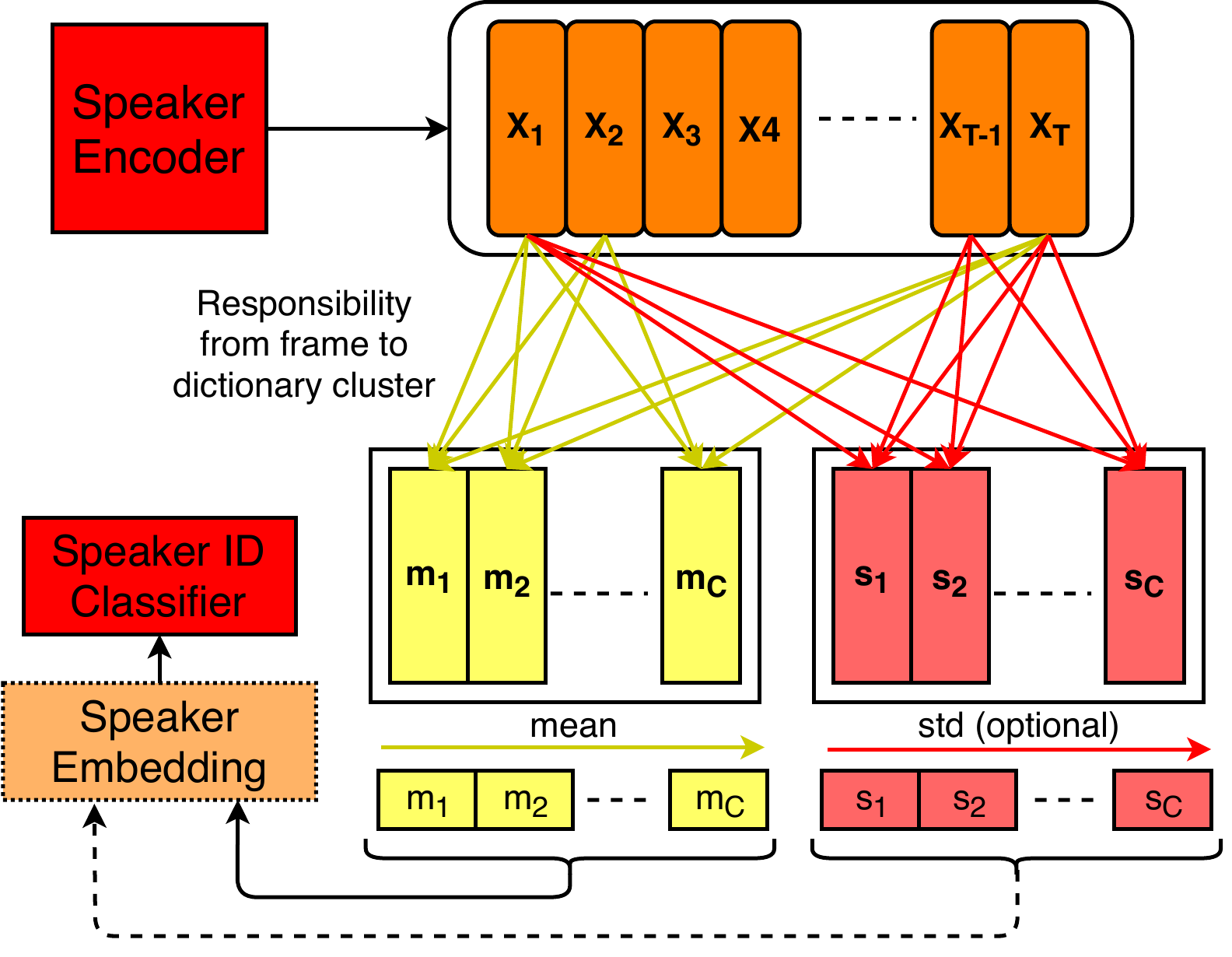}
    \vspace{-3mm}
    \caption{Learnable dictionary encoding (LDE) pooling method in an end-to-end speaker recognition system.}
    \label{fig:lde_fig}
    \vspace{-5mm}
\end{figure}

Given the frame-level representations $\mathbf{x}_{T} = \{x_{1}, x_{2}, ..., x_{T}\}$ from the encoder networks, where $T$ is the sequence length, an LDE layer learns a dictionary of $C$ clusters $\{e_{1}, e_{2}, ..., e_{C}\}$. The learning procedure is decomposed into three steps: 1) compute some distance $r_{tc}$ from each frame $x_{t}$ to each cluster $e_{c}$, 2) learn a soft cluster weight $w_{tc}$ of $x_{t}$ to $e_{c}$ based on $r_{tc}$, and 3) aggregate $x_{t}$ based on $w_{tc}$ over time $T$ to yield an utterance-level representation. Here $r_{tc}$ is L2 distance, $r_{tc} = \norm{x_{t} - e_{c}}^2$. The cluster weight $w_{tc}$ can be computed by, 
$w_{tc} = \exp(-r_{tc})/\sum_{i=1}^{C} \exp(-r_{ti})$. 
The aggregation of $x_{t}$ is similar to the supervector notion in \cite{reynolds2000speaker}. 
We first compute mean $m_{c} = \frac{1}{Z}\sum_{\forall t} w_{tc}(x_{t} - e_{c})$ and/or standard deviation  $s_{c} = \frac{1}{Z} \sqrt{\sum_{\forall t} w_{tc}(x_{t} - e_{c})^2}$ for each cluster, which are concatenated for $\forall c \in \{1..C\}$ to form a mean vector $\mathbf m_C$, and similarly for a standard deviation vector $\mathbf s_C$. 
Here $Z = \sum_{t=1}^T w_{tc}$. Figure \ref{fig:lde_fig} illustrates these three steps. 
        
\vspace{-2mm}        
\subsection{Classifier}

As shown in Figure \ref{fig:lde_fig}, the last step is to predict speaker IDs via the softmax layer. The standard training criterion is therefore cross entropy. More discriminative criteria called angular softmax loss (A-softmax) and their variants have recently been proposed and evaluated in \cite{Huang2018,xiang2019margin, bhattacharya2019deep}. The criteria considers angular margins between classes and is expected to produce more separable embedding representations. We examine both normal softmax and angular softmax. 
  
\vspace{-2mm}        
\section{Multi-Speaker TTS Model Architecture}
\label{sec:pagestyle}

\begin{figure}[tb] \centering
\includegraphics[width=0.9\columnwidth]{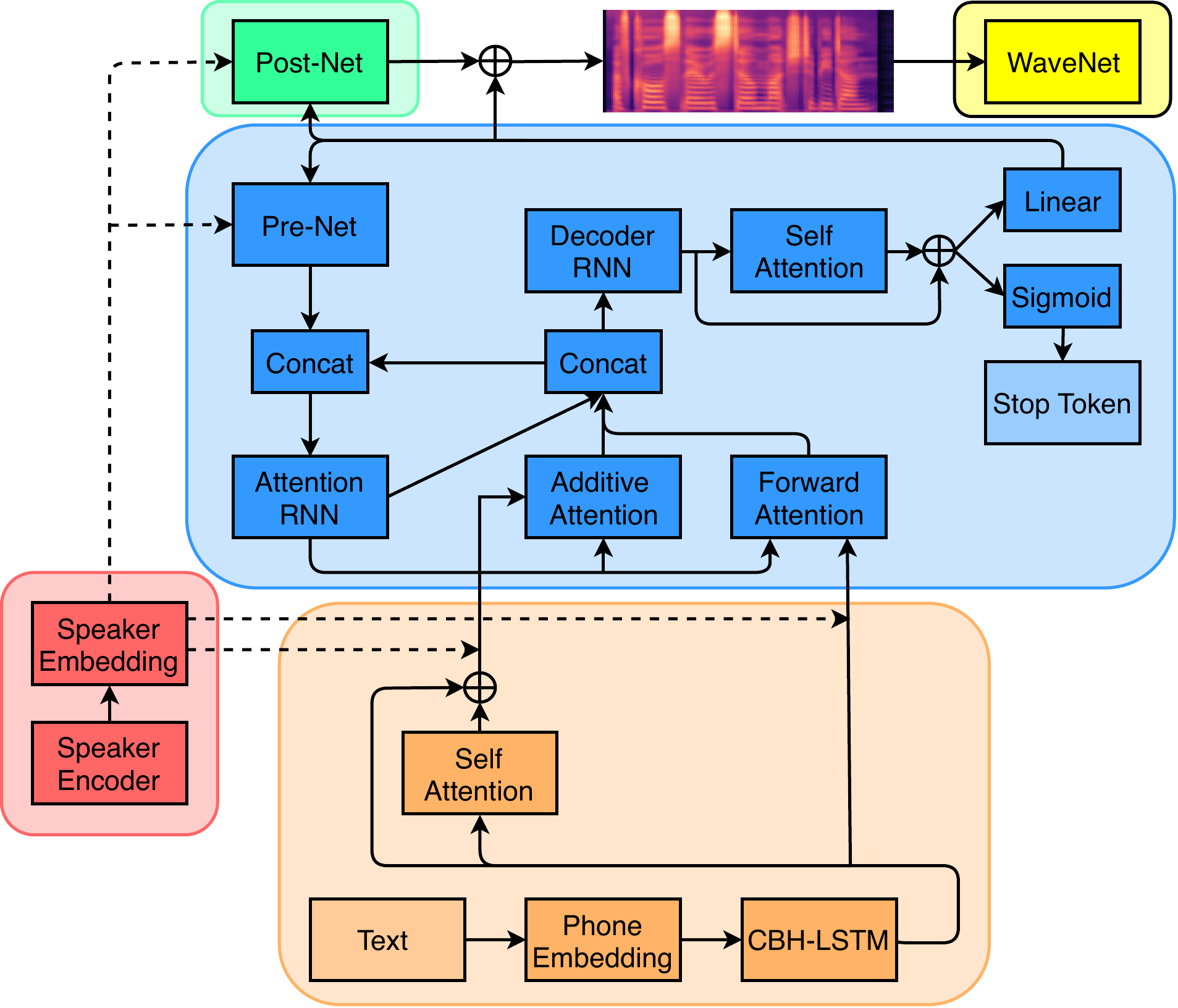}
\vspace{-3mm}
\caption{Proposed multi-speaker TTS system. Encoder blocks are in \textit{orange}, decoder blocks in \textit{blue}, post-net block in \textit{green}, speaker encoder blocks in \textit{red}, and vocoder block in \textit{yellow}.}
\label{fig:tts_fig}
\vspace{-5mm}
\end{figure}

The above speaker embedding vectors are used as additional inputs to condition speaker characteristics in our multi-speaker TTS system. 
Our end-to-end multi-speaker text-to-speech model architecture is based on Tacotron \cite{wang2017tacotron}, with the extension of self-attention described in \cite{yasuda2019investigation} to better capture long-range dependencies illustrated in Figure \ref{fig:tts_fig}.  We use phoneme input. We carry out basic rule-based text normalization to expand abbreviations and numbers.  We then convert the text to a phoneme representation using flite \cite{flite}. A self-attention block \cite{Vaswani2017} is added in the encoder, and so the encoder produces two outputs: one is the original output of the long short-term memory (LSTM), and the other is the output from the self-attention block.  The LSTM output is passed to a forward attention block \cite{Zhang2018}, which speeds up the alignment, and the self-attention output is passed to an additive attention block, which allows attention to longer-range information.  The outputs of the dual attention mechanism are concatenated before being passed to the decoder. The output of the decoder is an 80-dimensional mel-spectrogram. 

We consider three possible locations to input speaker embeddings: concatenating with each of the encoder outputs before inputting to the attention mechanisms, inputting to the prenet to the decoder, or inputting to the postnet. We extract a speaker embedding vector from each training utterance using the speaker encoder and average them per speaker. We then project all speaker embeddings down to 64 dimensions using a dense layer before inputting them to any location in the model.

Speaker adaptation to new speakers is zero-shot. As in the training phase, we extract a speaker embedding vector from each untranscribed adaptation utterance of a target speaker using the speaker encoder. We then input the averaged speaker embedding to generate mel spectrograms of the target speaker. No fine tuning is used. To convert the predicted mel spectrograms into audio, we use a WaveNet \cite{wavenet} vocoder.  Input is mel spectrograms and output is 16-bit 16kHz waveforms.  The code for our multi-speaker Tacotron implementation and audio samples will be available online\footnote{\href{https://github.com/nii-yamagishilab/multi-speaker-tacotron}{https://github.com/nii-yamagishilab/multi-speaker-tacotron}}\footnote{\href{https://nii-yamagishilab.github.io/samples-multi-speaker-tacotron}{https://nii-yamagishilab.github.io/samples-multi-speaker-tacotron}}.

\vspace{-2mm}        
\section{Experiments}
\label{sec:exp}
\vspace{-2mm}
\subsection{Speaker Verification}
    \label{sec:svexp}
    In the following sections, we will refer to speaker embeddings based on TDNN+SP as the x-vectors and those based on ResNet34+LDE as the LDEs. \\
        \noindent\textbf{Data}:
        Following \cite{jia2018transfer}, we trained speaker verification systems on VoxCeleb1+2 \cite{nagrani2017voxceleb, chung2018voxceleb2}. The training data were all of VoxCeleb2 plus the training portion of VoxCeleb1 (7,325 speakers and 1,277,344 utterances). The clean speech was augmented with reverberation, noise, music, and babble, as described in \cite{snyder2018x}, and then a random subset of these 1,000,000 augmented utterances was combined with the original clean speech. The final training data  consisted of 2,277,344 utterances. We report the speaker verification results on the original VoxCeleb1 test set.
        
        \noindent\textbf{Acoustic Features and Pre-processing}:
        We trained x-vectors on 30-dimensional MFCCs and LDEs with 30-dimensional log-Mel filter banks. Kaldi-based 3-second sliding cepstral mean normalization and energy VAD were applied to the acoustic features. This is similar to the setup described in \cite{villalba2019state}. For the LDE systems, each training sample was a 3-8 second chunk randomly sampled from its original utterance. Our chunk length selection is consistent with \cite{cai2018exploring, chen2019tied}.
    
        \noindent\textbf{System Details}:
        Our x-vectors are based on the Kaldi recipe\footnote{\href{https://github.com/kaldi-asr/kaldi/tree/master/egs/voxceleb/v2}{https://github.com/kaldi-asr/kaldi/tree/master/egs/voxceleb/v2}}, with 512 dimensions. For the LDEs, our ResNet34 is the same as \cite{villalba2019state, cai2018exploring, chen2019tied}, and we set the number of dictionary clusters $C = 32$ and mini-batch size to 128. We experimented with the following hyperparameter combinations: embedding dimension $\{512, 256, 200\}$, softmax margin $m\in\{2, 3, 4\}$, and pooling only mean vector $\mathbf m_C$ or both mean and standard deviation vectors $\mathbf m_C$ and $\mathbf s_C$.
        
        \noindent\textbf{Embedding Post-processing and Backend}:
        Our backend is PLDA \cite{kenny2013plda} with score normalization. We followed Kaldi's backend recipe in post-processing the embeddings prior to the PLDA: centering and LDA reduction to 200 dimensions. We also scored the original embeddings without this post-processing step, as we were interested in the effect of such a procedure for speaker adaption in TTS. Note that we did not perform length normalization nor any adaptation/tuning, as normally done in speaker verification, such as \cite{villalba2019state}.
        
                \begin{table}[t!]
          \caption{\it Verification results on the original VoxCeleb1 test set. $\mathbf m, \mathbf s$ are mean $\mathbf m_C$ and standard deviation $\mathbf s_C$ vectors. S is softmax, AS is A-softmax and the number within parentheses is the angular margin $m$. \textit{norm} indicates post-processing (centering+LDA) on the embeddings. $C = 32$ is set for all LDEs. EER denotes equal error rate and DCF$_{0.01}^{min}$ denotes minimum detection cost function value with a prior value set to 0.01 \cite{BRUMMER2006230}.} \label{tab:spk_ver_table}
          \centering
          \scalebox{0.9}{
            \begin{tabular}{@{}|l||c|c|c|c|c|c|@{}}
            \hline
            \cellcolor{green!25}embed. & dim. & pl. & obj. & norm & EER & DCF$_{0.01}^{min}$ \\
            \hline
            \hline
            \cellcolor{green!25}i-Vec$^{\bf N}$ & 400 & $\mathbf m$ & EM & \checkmark & 5.329 & 0.493 \\
            \hline
            \hline
            \cellcolor{green!25}x-Vec & 512 & $\mathbf m, \mathbf s$ & S & & 3.298 & 0.343 \\
            \hline
            \cellcolor{green!25}x-Vec$^{\bf N}$ & 512 & $\mathbf m, \mathbf s$ & S & \checkmark & 3.213 & 0.342 \\
            \hline
            \hline
            \cellcolor{green!25}LDE-1 & 512 & $\mathbf m$ & S & & 3.415 & 0.366 \\
            \hline
            \cellcolor{green!25}LDE-1$^{\bf N}$ & 512 & $\mathbf m$ & S & \checkmark  & 3.446 & 0.365 \\
            
            \hline
            \cellcolor{green!25}LDE-2 & 512 & $\mathbf m$ & AS(2) & & 3.674 & 0.364 \\
            \hline
            \cellcolor{green!25}LDE-2$^{\bf N}$ & 512 & $\mathbf m$ & AS(2) & \checkmark & 3.664 & 0.386 \\
            
            \hline
            \cellcolor{green!25}LDE-3 & 512 & $\mathbf m$ & AS(3) & & \cellcolor{blue!25} \bf  3.033 & \cellcolor{red!25} \bf 0.314 \\
            \hline
            \cellcolor{green!25}LDE-3$^{\bf N}$& 512 & $\mathbf m$ & AS(3) & \checkmark & 3.171 & 0.327 \\
            
            \hline
            \cellcolor{green!25}LDE-4 & 512 & $\mathbf m$ & AS(4) & & 3.112 & 0.315 \\
            \hline
            \cellcolor{green!25}LDE-4$^{\bf N}$ & 512 & $\mathbf m$ & AS(4) & \checkmark & 3.271 & 0.327 \\
            
            \hline
            \cellcolor{green!25}LDE-5 & 256 & $\mathbf m$ & AS(2) & & 3.287 & 0.343 \\
            \hline
            \cellcolor{green!25}LDE-5$^{\bf N}$ & 256 & $\mathbf m$ & AS(2) & \checkmark & 3.367 & 0.351 \\
            
            \hline
            \cellcolor{green!25}LDE-6 & 200 & $\mathbf m$ & AS(2) & & 3.266 & 0.396 \\
            \hline
            \cellcolor{green!25}LDE-6$^{\bf N}$ & 200 & $\mathbf m$ & AS(2) &\checkmark & 3.266 & 0.396 \\
            
            \hline 
            \cellcolor{green!25}LDE-7 & 512 & $\mathbf m, \mathbf s$ & AS(2) & & \cellcolor{red!25} \bf 3.091 & \cellcolor{blue!25} \bf 0.303 \\
            \hline 
            \cellcolor{green!25}LDE-7$^{\bf N}$ & 512 & $\mathbf m, \mathbf s$ & AS(2) & \checkmark & 3.171 & 0.328 \\
            \hline
          \end{tabular}
          }
          \vspace{-5mm}
        \end{table}
        
        \noindent\textbf{Verification Results}:
        Table \ref{tab:spk_ver_table} presents the results of speaker verification on the VoxCeleb1 test set. We denoted our 7 LDE embeddings as LDE-1, LDE-2, etc, and used superscript a $\bf N$ to mark those with the post-processing step described above. The LDEs attained results on par with the x-vectors. We also observed that decreasing speaker embedding size, increasing angular margin $m$, and pooling both $\mathbf m_C$ and $\mathbf s_C$ improve the performance.

\vspace{-3mm}    
\subsection{Preliminary Experiments for Speaker Similarity in TTS}
    \vspace{-0.5mm}
    Since the best training method and location for inputting speaker embeddings to the TTS was unknown, we conducted  preliminary experiments to learn which settings produce the best speaker similarity for unseen speakers.  We wanted to learn whether it was better to train gender-dependent or gender-independent models, and whether it is best to input speaker embeddings to the prenet, concatenate with the encoder output, input at the postnet, or some combination of these.
    
    \noindent\textbf{Data:} We used the VCTK corups \cite{vctk}, which consists of read English speech from 109 different speakers in different English dialects.  Each speaker read about 400 sentences.  Two speakers were excluded due to missing or inadequate data.  Four development and four test speakers were held out, chosen to be a mix of genders and dialects, and to have enough unique utterances to have 50 unseen sentences per speaker for TTS evaluation and 50 unseen utterances for ``adaptation data'' for extracting speaker embeddings.  Audio was preprocessed by first high-pass filtering at a cutoff of 80 Hz to remove low-frequency line noise, then normalized using sv56 \cite{sv56}, then trimmed to remove start and end silences.  All utterances from the 99 training (``seen'') speakers were used to train TTS and to extract speaker embeddings for these speakers; this same data was used to train gender-dependent WaveNet vocoders.  The embeddings of the four development and four test speakers (``unseen'' speakers) were extracted using only the 50 held-out ``adaptation'' utterances.
    
    \noindent\textbf{Training:}  We used a ``warm-start''  training approach to reduce experimental iteration time.  We initialized our multi-speaker models with parameters from a well-trained speaker-dependent model trained on the ``Nancy'' data from Blizzard 2011 \cite{blizzard2011} for about 105k steps.  We experimentally found that multi-speaker models trained with warm-start for one day (about 40k steps) produced synthetic speech of similar quality to models trained from scratch on VCTK data only for four days.  Furthermore, \cite{taylor2019analysis} observed that the VCTK corpus has a relatively small number of unique words, whereas the Nancy dataset has more than three times as many; our multi-speaker model can benefit from this increased lexical coverage.

    \noindent\textbf{Settings:}  We tried a number of different settings to determine which had the best similarity for unseen speakers.  For speaker embeddings, we used x-vectors.  We tried two different training approaches:
        \begin{itemize}
            \item \textbf{Gender-independent:} We used data from all VCTK training speakers (male and female) for warm-start training.
            \item \textbf{Gender-dependent:} We ran two separate warm-start trainings, one using only male VCTK training data and the other using only the female data.
        \end{itemize}
    At the same time, we tried four different settings for the location to input speaker embeddings:
        \begin{itemize}
            \item Prenet only (\texttt{pre})
            \item Concatenate with encoder output only and input to attention mechanism (\texttt{attn})
            \item Prenet + concatenate with encoder output (\texttt{pre+attn})
            \item Prenet + concatenate with encoder output + postnet \\
            (\texttt{pre+attn+post})
        \end{itemize}
We did not try postnet input alone because we found that this configuration produced poor quality synthetic speech, but we decided to investigate its combination with other input locations.

    \noindent\textbf{Evaluation and Results:}  
    We objectively evaluated the different combinations of training strategy and embedding input locations by synthesizing some sample utterances from four ``seen'' speakers (those included in training) and four ``unseen'' ones (development speakers).  Since we did not hold out any data from the ``seen'' speakers' utterances, we synthesized seen speakers' sample utterances from a randomly selected set of texts from the test set (unseen during training).  We then extracted x-vectors for each speaker from the {\em synthesized} speech, and measured cosine similarity to the target speaker's x-vector extracted from his or her actual speech.  Cosine similarity is defined as $cos\_sim(A,B) = A \cdot B/\norm{A}\norm{B}$ and is a standard measure of similarity of speaker embedding vectors for ASV.  The values range from -1 to 1, and higher values indicate that the vectors are more similar.  Cosine similarity results for different configurations are listed in Table \ref{tab:cossim}.
    
            \begin{table}[t!]
          \centering
          \footnotesize
         \caption{\it Average cosine similarities between original and synthesized speech from different model configurations for seen (training) and unseen (dev set) speakers. Waveform generation was done using unseen texts for both seen and unseen speakers.}
        \label{tab:cossim}
            \begin{tabular}{@{}|l||c|c||c|c|@{}}
            \hline 
            Input location & \multicolumn{2}{c||}{Gender-ind} & \multicolumn{2}{c|}{Gender-dep}  \\
            \hline
                  & train & dev & train & dev \\
            \hline \hline
            pre & 0.357 & 0.402 & 0.438 & 0.361 \\
            \hline
            attn & 0.709 & 0.490 & 0.711 & 0.476 \\
            \hline
            pre+attn & 0.676 & 0.489 & 0.708 & \cellcolor{blue!25} \textbf{0.533} \\
            \hline
            pre+attn+post & 0.684 & 0.480 & 0.717 & 0.477 \\
            \hline 
          \end{tabular}
          \vspace{-5mm}
        \end{table}

    As expected, we see a gap between seen and unseen speakers: seen speakers' synthetic speech generally has higher similarity to the original speech.    
    Since the gender-dependent training with x-vectors input at both the prenet and attention mechanism produced the synthetic speech with best similarity for unseen speakers, we chose this configuration for our later experiments.

\vspace{-2mm}    
\subsection{Comparing Different Embeddings for Speaker Similarity}
    
    After we chose the best training and model settings (gender-dependent training with embedding input at both the prenet and attention mechanism), we trained 15 TTS models each using a different type of speaker embedding: the 14 types of LDE embeddings described in Section \ref{sec:svexp} and x-vectors.  We then conducted a crowdsourced listening test to evaluate  naturalness and speaker similarity for both seen and unseen speakers using each speaker embedding as well as copy-synthesized speech and natural speech for comparison.  
    
    For each TTS system, we synthesized 50 sentences from each of the four ``seen'' (training) and eight ``unseen'' (development and test) speakers, for a total of 600 unseen test utterances per system.  Listeners heard one test utterance at a time and first rated it on a Likert scale from 1-5 for Mean Opinion Score (MOS) for naturalness, then rated for speaker similarity compared to a reference utterance on a Differential MOS (DMOS) scale \cite{Lorenzo-Trueba2018} from 1 (definitely a different speaker) to 4 (definitely the same speaker).  Reference utterances were randomly chosen from the target speaker's original speech.  Listeners rated ``sets'' of 25 utterances, and each listener could complete a maximum of ten sets.  Each set was completed by five different listeners, and 463 subjects participated.  Sets were designed to contain at least one utterance from every system to average out listener differences over all systems.  Results are in Table \ref{tab:mos}.  Natural speech was rated with MOS of 3.83 and DMOS of 3.25.
    
    We found that speaker similarity scores for speakers seen during training are very close to those for vocoded speech.  Similarity scores for unseen speakers (dev and test) are also lower than seen speakers, as expected, and consistent with Table \ref{tab:cossim}.  We observed that advanced neural speaker embeddings improve speaker similarity for unseen speakers compared to x-vectors.  Unexpectedly, they also improve naturalness.  While LDE helps, the impact of angular softmax and postprocessing (N) seems to be small.  
    For completely unseen test set speakers, system LDE-3 was best in terms of both naturalness and speaker similarity.  This system was significantly better than the x-vector system according to a Mann-Whitney U test both in terms of naturalness (p=5.9e-11) and speaker similarity (p=0.02).  This was also the best type of embedding in terms of EER.  We did not find any meaningful correlations between ASV and TTS scores.

        \begin{table}[t!]
          \caption{\it MOS and DMOS results for seen (train) and unseen (dev and eval) speakers using each type of speaker embedding. Waveform generation was done using unseen texts for all speakers. Five-point and four-point scales were used for naturalness and similarity evaluation, respectively.  Blue boxes show the best results for each condition and red boxes show second and third best.} \label{tab:mos}
          \vspace{2mm}
          \centering
          \scalebox{0.9}{
            \begin{tabular}{@{}|l||c|c|c||c|c|c|@{}}
            \hline
             & \multicolumn{3}{c||}{Naturalness} & \multicolumn{3}{c|}{Similarity}  \\
            \hline
            \cellcolor{green!25}system & train & dev & test & train & dev & test \\
            \hline
            \hline
            \cellcolor{green!25}vocoded & 3.51  & 3.41  & 3.55  & 3.02 & 2.79  & 2.82 \\
            \hline
            \hline
            \cellcolor{green!25}x-Vec$^{\bf N}$ & 3.20 & 3.19 & 3.19 & 2.93 & 1.86 & 2.37  \\
            \hline
            \hline
            \cellcolor{green!25}LDE-1 & 3.15 & 3.16 & 3.21 & 2.87 & \cellcolor{blue!25} \bf 2.05 & 2.34  \\
            \hline
            \cellcolor{green!25}LDE-1$^{\bf N}$ & 3.04 & 3.13 & \cellcolor{red!25} \bf 3.46 & 2.87 & 1.97 & \cellcolor{red!25} \bf 2.45  \\
            
            \hline
            \cellcolor{green!25}LDE-2 & 3.11 & \cellcolor{red!25} \bf3.28 & 3.35 & 2.84 & 2.00 & 2.37  \\
            \hline
            \cellcolor{green!25}LDE-2$^{\bf N}$ &  3.13 & 3.19 & 3.33 & 2.90 & 2.00 & 2.35  \\
            
            \hline
            \cellcolor{green!25}LDE-3 & 3.09 & 3.24 & \cellcolor{blue!25} \bf 3.48 & 2.89 & 1.88 & \cellcolor{blue!25} \bf 2.46  \\
            \hline
            \cellcolor{green!25}LDE-3$^{\bf N}$& 3.14 & 3.16 & 3.33 & 2.91 & 2.00 & 2.37  \\
            
            \hline
            \cellcolor{green!25}LDE-4 & 3.08 & 3.10 & 3.29 & 2.94 & 2.00 & 2.31  \\
            \hline
            \cellcolor{green!25}LDE-4$^{\bf N}$ & 3.12 & 3.20 & 3.29 & 2.90 & 1.98 & 2.39  \\
            
            \hline
            \cellcolor{green!25}LDE-5 & 3.07 & \cellcolor{red!25} \bf3.26 & \cellcolor{red!25} \bf 3.40 & 2.89 & 1.99 & \cellcolor{red!25} \bf 2.45  \\
            \hline
            \cellcolor{green!25}LDE-5$^{\bf N}$ & 3.11 & 3.07 & 3.37 & 2.88 & \cellcolor{red!25} \bf 2.02 & 2.41  \\
            
            \hline
            \cellcolor{green!25}LDE-6 & 3.12 & 3.25 & 3.33 & 2.92 & 1.95 & 2.43  \\
            \hline
            \cellcolor{green!25}LDE-6$^{\bf N}$ & 3.13 & \cellcolor{blue!25} \bf 3.29 & 3.23 & 2.88 & 1.94 & 2.39  \\
            
            \hline 
            \cellcolor{green!25}LDE-7 & 3.15 & 3.03 & 3.18 & 2.91 & 1.86 & 2.28  \\
            \hline 
            \cellcolor{green!25}LDE-7$^{\bf N}$ & 3.07 & 3.02 & 3.24 & 2.83 & \cellcolor{red!25} \bf 2.02 & 2.42  \\
            \hline
          \end{tabular}
          }
          \vspace{-5mm}
        \end{table}

\section{Conclusions}
\label{sec:discussion}
We found that the LDE-based neural speaker embeddings can improve speaker similarity and naturalness of synthetic speech for unseen speakers, and this approach can be used for zero-shot speaker adaptation.  However, there is still a gap between seen and unseen speaker similarity, indicating that the TTS model may still be overfitting to seen speakers and there is room for improvement.  For future work, we will explore ways to mitigate this overfitting by trying different methods of speaker space augmentation.  We would also like to evaluate adaptation performance on more nuanced aspects of speaker similarity, such as dialect and speaking style.

\vspace{3mm}

\noindent
\textbf{Acknowledgments}
\footnotesize{This work was partially supported by a JST CREST Grant (JPMJCR18A6, VoicePersonae project), Japan, and by MEXT KAKENHI Grants (16H06302, 17H04687, 18H04120, 18H04112, 18KT0051, 19K24372), Japan. The numerical calculations were carried out on the TSUBAME 3.0 supercomputer at the Tokyo Institute of Technology.}

\vfill\pagebreak

\bibliographystyle{IEEEtran}
\bibliography{main}

\end{document}